\documentclass[%
 reprint,
 superscriptaddress,
 amsmath,amssymb,
 aps,
 prl,
floatfix,
]{revtex4-2}

\usepackage{graphicx}
\usepackage{dcolumn}
\usepackage{bm}
\usepackage{physics}

\begin{document}

\title{Element-specific probe of quantum criticality in CeCoIn$_5$}

\author{A. Khansili}
\email{Authors contributed equally}
\affiliation{School of Materials Science and Technology, Indian Institute of Technology (Banaras Hindu University), Varanasi -  221005, India}%
\affiliation{Department of Physics, Stockholm University, SE-106 91 Stockholm, Sweden}

\author{R. Sharma}
\email{Authors contributed equally}
\affiliation{School of Materials Science and Technology, Indian Institute of Technology (Banaras Hindu University), Varanasi -  221005, India}%
 
\author{R. Hissariya}
\affiliation{School of Materials Science and Technology, Indian Institute of Technology (Banaras Hindu University), Varanasi -  221005, India}%

\author{I. Baev}
\affiliation{Universit{\"{a}}t Hamburg, Institut f{\"{u}}r Experimentalphysik Luruper Chaussee 149, Hamburg, Germany}%

\author{J. Schwarz}
\affiliation{Universit{\"{a}}t Hamburg, Institut f{\"{u}}r Experimentalphysik Luruper Chaussee 149, Hamburg, Germany}%

\author{F. Kielgast}
\affiliation{Universit{\"{a}}t Hamburg, Institut f{\"{u}}r Experimentalphysik Luruper Chaussee 149, Hamburg, Germany}%

\author{M. Nissen}
\affiliation{Universit{\"{a}}t Hamburg, Institut f{\"{u}}r Experimentalphysik Luruper Chaussee 149, Hamburg, Germany}%

\author{M. Martins}
\affiliation{Universit{\"{a}}t Hamburg, Institut f{\"{u}}r Experimentalphysik Luruper Chaussee 149, Hamburg, Germany}%

\author{M. -J. Huang}
\affiliation{Deutsches Elektronen-Sychrotron DESY, Notkestraße 85, 22607 Hamburg, Germany}%

\author{M. Hoesch}
\affiliation{Deutsches Elektronen-Sychrotron DESY, Notkestraße 85, 22607 Hamburg, Germany}%

\author{V. K. Paidi}
\affiliation{Pohang Accelerator Laboratory, Pohang 37673, South Korea}%
\affiliation{Department of Physics and Astronomy, University of Manitoba, Winnipeg, Manitoba  R3T 2N2, Canada}%

\author{J. van Lierop}
\affiliation{Department of Physics and Astronomy, University of Manitoba, Winnipeg, Manitoba  R3T 2N2, Canada}%

\author{A. Rydh}
\email{andreas.rydh@fysik.su.se}
\affiliation{Department of Physics, Stockholm University, SE-106 91 Stockholm, Sweden}

\author{S. K. Mishra}
\email{shrawan.mst@iitbhu.ac.in}
\affiliation{School of Materials Science and Technology, Indian Institute of Technology (Banaras Hindu University), Varanasi -  221005, India}
 
\date{\today}

\begin{abstract}
Employing the elemental sensitivity of x-ray absorption spectroscopy (XAS) and x-ray magnetic circular dichroism (XMCD), we study the valence and magnetic order in the heavy fermion superconductor CeCoIn$_5$. We probe spin population of the f-electrons in Ce and d-electrons in Co as a function of temperature (down to $0.1$\,K) and magnetic field (up to $6$\,T). From the XAS we find a pronounced contribution of Ce$^{4+}$ component at low temperature and a clear temperature dependence of the Ce valence below $5$\,K, suggesting enhanced valence fluctuations, an indication for the presence of a nearby quantum critical point (QCP). We observe no significant corresponding change with magnetic field.
The XMCD displays a weak signal for Ce becoming clear only at $6$\,T. This splitting of the Kramers doublet ground state of Ce$^{3+}$ is significantly smaller than expected for independent but screened ions, indicating strong antiferromagnetic pair interactions. The unconventional character of superconductivity in CeCoIn$_5$ is evident in the extremely large specific heat step at the superconducting transition. 

\begin{description}
\item[Keywords]
Unconventional superconductivity, quantum criticality, strongly correlated electrons, quantum phase transition, x-ray absorption spectroscopy, x-ray magnetic circular dichroism    
\end{description}
\end{abstract}

\maketitle

Unconventional superconductivity has long been a fascinating topic in condensed matter physics. Its microscopic origin however remains elusive \cite{scalapino2012common,norman2011challenge}.
Manifestations of unconventional superconductivity extend over a wide range of materials, including cuprates \cite{tsuei2000pairing}, iron pnictides or chalcogenides \cite{dai2015antiferromagnetic}, topological insulators \cite{linder2010unconventional}, and heavy fermions \cite{gegenwart2008quantum}.
For materials containing f-orbitals, the interaction between f-electron spins and itinerant conduction electrons can lead to low-energy quasiparticles of heavy effective mass.
Exceptional magnetic properties may display in heavy-fermion systems in the vicinity of a magnetic quantum critical point (QCP)\cite{scalapino2012common} due to the interaction of conduction electrons, localized moments, and the competition between potential ground states.
The interaction with localized magnetic moments induced by the rare-earth magnetic ion transforms the physical properties by generating quasiparticles that show distinctive properties from their constituents and related non-Fermi-liquid (NFL) behavior \cite{stewart2001non}.
QCPs may be accompanied by changes in the Fermi surface, enhanced quantum fluctuations, and may sometimes result in superconductivity or other novel ground states \cite{petrovic2001heavy,sidorov2002superconductivity,maksimovic2022evidence}.

The CeMIn$_5$ (M=Co, Ir, and Rh), i.e., Ce-115 family is one of the most studied heavy fermion superconductors \cite{sidorov2002superconductivity,maksimovic2022evidence,petrovic2001new}. These compounds are f-electron systems with a crystal structure that resembles that of the high-$T_c$ cuprates. Both Ce-115 and cuprates display a non-thermal parameter such as doping- or pressure-driven quantum phase transition (QPT) underlying the superconducting state. Among Ce-based heavy fermion compounds, the Ce-115 family is one of the most intriguing due to its high $T_c$, with the highest $T_c$ of 2.3\,K occurring for CeCoIn$_5$ located intrinsically near a QCP \cite{pham2006reversible}. Over the last three decades, CeCoIn$_5$ has emerged as an archetypical system to investigate the characteristics of QCP \cite{petrovic2001heavy,sidorov2002superconductivity,maksimovic2022evidence}.

So far, no low-temperature element-specific study is available to characterize the quantum criticality for the Ce-115 family. Here we study the low-temperature f-electron valence state in Ce and underlying magnetic order of CeCoIn$_5$ through  element-specific x-ray absorption spectroscopy (XAS) and x-ray magnetic circular dichroism (XMCD) down to 0.1\,K. The results support the view that CeCoIn$_5$ is imminent to a delocalization quantum phase transition with change of f-electron orbital occupancy becoming pronounced below 5\,K. 

We studied single crystals of CeCoIn$_5$ grown through a flux-melt method using excess indium as flux. The phase purity of crystals was probed by energy-dispersive x-ray spectroscopy (EDX) technique. The EDX pattern (supplementary Fig.~S1) confirms single-phase CeCoIn$_5$. The XAS spectra were measured at P04, PETRA III, DESY (Hamburg), in the total electron yield (TEY) mode inside an ultra-high vacuum (UHV) chamber with a pressure in the 10$^{-10}$\,mbar range. Undulator P04 beamline enables us to switch the circular polarization of incoming x-rays. A clean surface was used for the M$_{5,4}$ edge of the XAS and XMCD measurements. The sample was mounted at a grazing angle of $25^{\circ}$ and the TEY signal was normalized to the incoming photon flux $I_{0}$, which was collected using a gold mesh. External magnetic field was applied through a superconducting magnet and always parallel to the incoming beam ($25^{\circ}$ to the basal $ab$-plane) \cite{beeck2016new}. 

CeCoIn$_5$ is derived from cubic CeIn$_3$ intercalated with CoIn$_2$ layers in a tetragonal structure (space group P4/mmm) with a unit cell $a\,=\,b\,=\,4.65$\,Å, $c\,=\,7.54$\,Å \cite{moshopoulou2002comparison}. The parent compound CeIn$_3$ forms long-range antiferromagnetic order at $T_\mathrm{N} =10$\,K. When $T_\mathrm{N}$ is suppressed by applied pressure, it becomes superconducting with $T_c = 0.2$\,K  \cite{walker1997normal}. CeCoIn$_5$ has a superconducting critical temperature $T_c = 2.3$\,K, but does not display long-range magnetic order \cite{ramos2010superconducting, yokoyama2008change}. The naively expected free magnetic moment $\mu = 2.54\,\mu_B$ and $\mu = 5.67\,\mu_B$, for Ce$^{3+}$ and Co$^{2+}$ respectively, are quenched due to crystal field (CF) effects. In CeCoIn$_5$ the interactions between Ce ions and with the conduction electrons are of antiferromagnetic (AFM) character \cite{yokoyama2008change,kenzelmann2008coupled}. The $T_c$ of the Ce-115 compounds varies linearly with the $c/a$ ratio of the tetragonal lattice parameters, highlighting the significance of the anisotropic electronic structure contributions \cite{bauer2004structural}. This variation might also be related to the CF, which displays anisotropic characteristics.

In CeCoIn$_5$ the Hund's rule (isotropic) ground state of the partially filled $4\mathrm{f}$-orbitals of Ce$^{3+}$ is $J$\,=\,5/2. The tetragonal crystal symmetry (point group $D_\mathrm{4h}$) causes CF splitting of this ground state into three Kramers doublets \cite{willers2010crystal}: 
\begin{alignat}{2}
&\ket{2} &&= \ket{\pm1/2}, \nonumber \\
&\ket{1} &&= \beta \ket{\pm5/2} - \alpha \ket{\mp 3/2}, \\
&\ket{0} &&= \alpha \ket{\pm5/2} + \beta \ket{\mp 3/2}, \nonumber
\end{alignat}
with ground state $\ket{0}$ and ${|\alpha|}^2+{|\beta|}^2=1$. From comparing XAS experimental results with simulations, the mixing factor has been found to be ${|\alpha|}^2 = 0.10$ \cite{willers2010crystal, sundermann2019orientation}. This rather small number is consistent with symmetry favoring $J_z = \pm 3/2$ states. With $\beta = 1$, the magnetic moment would be $1.0\,\mu_B$, but the mixing with $\pm 5/2$ reduces the moment further to $\mu \approx 0.7 \mu_B$. This reduction of Ce moment from 2.54\,$\mu_B$ to 0.7\,$\mu_B$ is due to CF effects alone.

\begin{figure}
\includegraphics[width=0.98\linewidth]{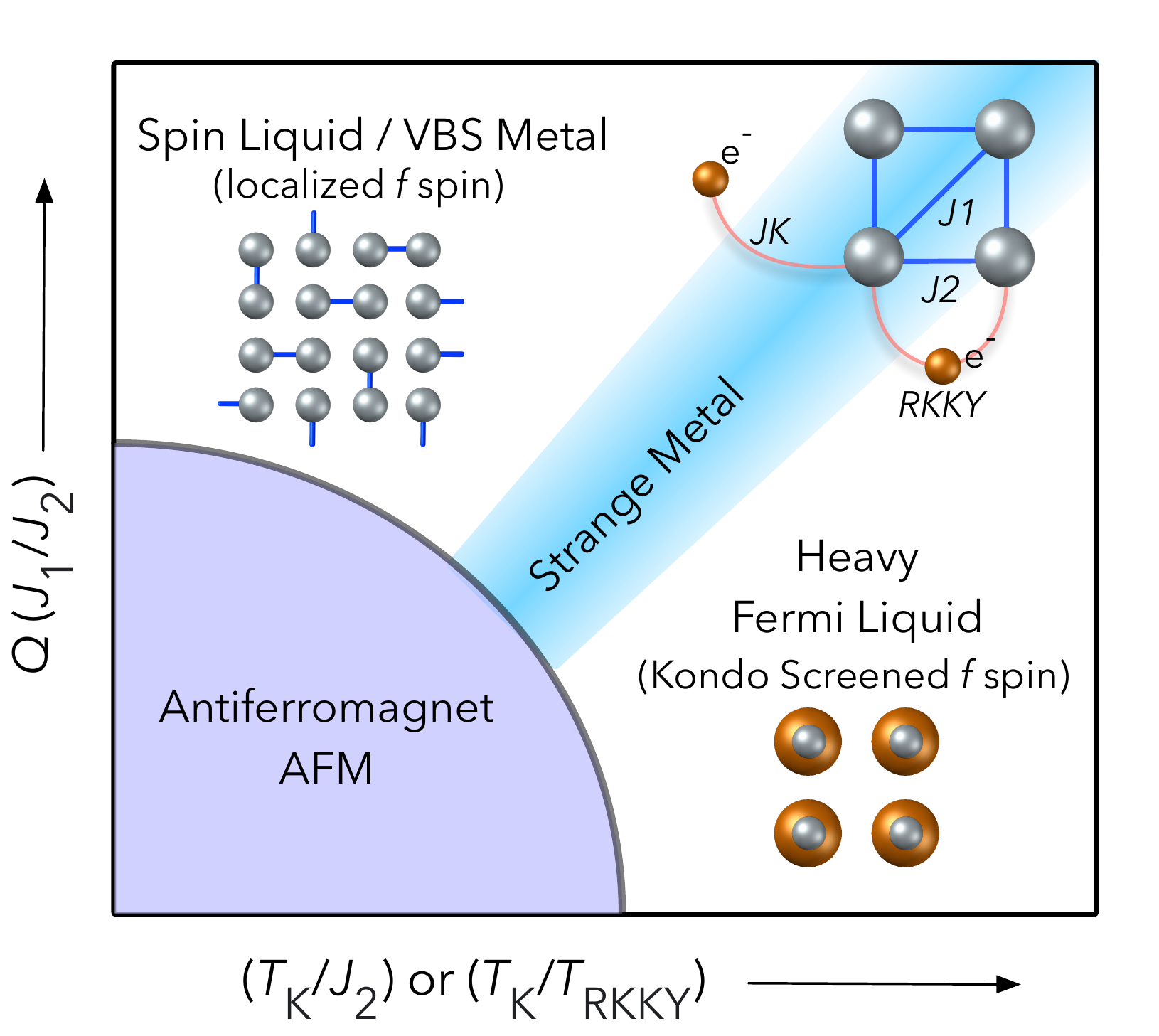}
\caption{\label{Fig1_QCP_phase_diagram} Zero-temperature phase diagram for heavy electron systems, illustrating ground state as a function of relative kondo strength and quantum frustration $Q$, in turn controlled by nearest-neighbour interaction $J_2$, next-nearest-neighbour interaction $J_1$, Kondo coupling $J_\mathrm{K}$ and RKKY interaction. These interaction are illustrates on the top right of the figure. A spin liquid or valence bond solid (VBS) metal is developed at high $Q$ and small $T_\mathrm{K}$. On the other hand, having a large $T_\mathrm{K}$ and small quantum frustration $Q$ will result in a heavy Fermi liquid with Kondo screened f-spins. AFM order develops when $J_2$ or RKKY  dominates. Strange metal phenomena occurs when these interactions have similar strength and there is a competition for the ground state of the system.}
\end{figure}

Magnetic materials usually attain an ordered state at the Curie/Néel temperature. However, despite having ions that can carry non-zero local magnetic moments, some of the heavy-fermion compounds lack long-range magnetic order even at low temperatures \cite{amusia2014theory,scalapino2012common,gegenwart2008quantum,sidorov2002superconductivity,petrovic2001heavy,maksimovic2022evidence,petrovic2001new,kenzelmann2008coupled}.
The \emph{system} ground state of such materials can be described in terms of the nearest-neighbour interaction\,($J_2$), next-nearest-neighbour interaction\,($J_1$) and Kondo coupling\,($J_\mathrm{K}$) \cite{coleman2010frustration}. The Kondo coupling can induce two different interaction mechanism governed by different energy scale ($T_\mathrm{K}$ and $T_\mathrm{RKKY}$) as seen in the doniach picture \cite{doniach1977kondo}. The interplay of these interactions is described schematically by the phase diagram in Fig.~\ref{Fig1_QCP_phase_diagram}.
Enhancement of the Kondo interaction ($T_\mathrm{K}$) over the RKKY interaction ($T_\mathrm{RKKY}$) and/or nearest-neighbour interaction\,($J_2$) gives rise to a heavy-fermion state with cooling. Eventually, coherent scattering of conduction electrons by $\mathrm{f}$-electrons develops \cite{coleman2015heavy}. CeCoIn$_5$ lies somewhere near the strange metal region, close to the AFM quantum phase transition as seen in doping studies \cite{pham2006reversible, bauer2005superconductivity}. Since we have multiple competing ground states in CeCoIn$_5$, one could expect correlation between quantum critical fluctuations and prevalence of Ce$^{4+}$ at low temperatures.

\begin{figure}
\includegraphics[width=0.98\linewidth]{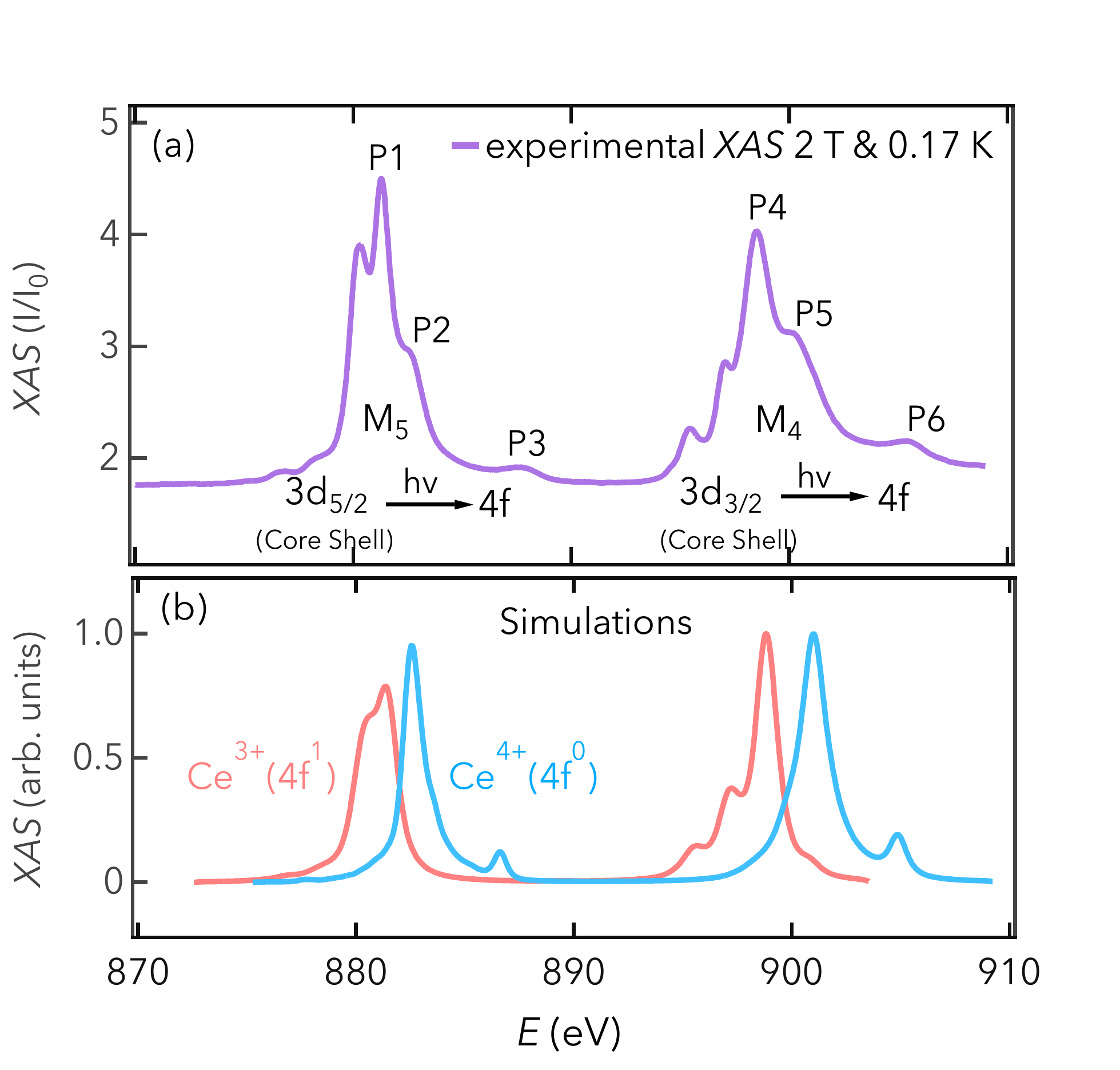}
\caption{\label{Fig2_XAS}(a) Ce-XAS normalized by the incoming light intensity ($I_0$) acquired at 0.17\,K under an applied external magnetic field of  2\,T. The spectrum clearly shows observation of M$_{5,4}$ resonance edges of Ce along with various satellite features (P2, P3, P5, and P6). (b) Simulated XAS for Ce$^{3+}$/Ce$^{4+}$ using the CTM4XAS software \cite{paidi2019role}.}
\end{figure}

The sensitivity of XAS to the valence state makes it a perfect tool to monitor the possible change of the $4\mathrm{f}$ occupation number $n_\mathrm{f}$ with external parameters. Such measurements at low temperature and high magnetic field, thus, enables the study of the valence state of the material and its proximity to a QCP. Figure~\ref{Fig2_XAS}\,(a) shows Ce-XAS spectra consisting of two main peaks at 881.3\,eV (P1 at M$_{5}$ edge) and 898.5\,eV (P4 at M$_{4}$ edge) along with weaker satellite peaks at 882.4\,(P2), 887.75\,(P3), 900\,(P5) and 905.5\,eV\,(P6).
The energy splitting between M$_{5,4}$ edges is due to the spin-orbit coupling of the 3d$_{5/2}$ and 3d$_{3/2}$ core electrons. The primary features of the M$_{5,4}$ edge XAS spectra originate from electric-dipole allowed transitions from $[3\mathrm{d}^{10}]\ldots 4\mathrm{f}^{n} \rightarrow [3\mathrm{d}^{9}]\ldots 4\mathrm{f}^{n+1}$ \cite{thole1985strong,paidi2019role,howald2015evidence,antonov2008x}. The XAS line shape depends strongly on the multiplet structures (P1-P6) as shown in Fig.~\ref{Fig2_XAS} (a). These multiplets represent the $3\mathrm{d}-4\mathrm{f}$ transition probabilites, affected by the $4\mathrm{f}$ state. Unique to soft x-ray absorption is that the dipole selection rules are very effective in determining which of the $4\mathrm{f}^{n+1}$ final states that can be reached and with which particular intensity.
From the XAS spectrum, information on the valence \cite{yamaoka2014role}, exchange and Coulomb interactions, local crystal fields \cite{willers2010crystal}, and hybridization can be obtained. Thus, this M$_{5,4}$-edge spectroscopy is extremely sensitive to the symmetry of the $4\mathrm{f}^{n}$ orbitals, especially the particular magnetic state of Ce$^{3+}$ ($n=1$) \cite{thole1985strong,paidi2019role,howald2015evidence,antonov2008x}. The spectral shapes of the Ce M$_{5,4}$ edges in Fig.~\ref{Fig2_XAS} (a) are indicative of Ce $4\mathrm{f}$-electrons being strongly hybridized.
\begin{figure}
\includegraphics[width=0.98\linewidth]{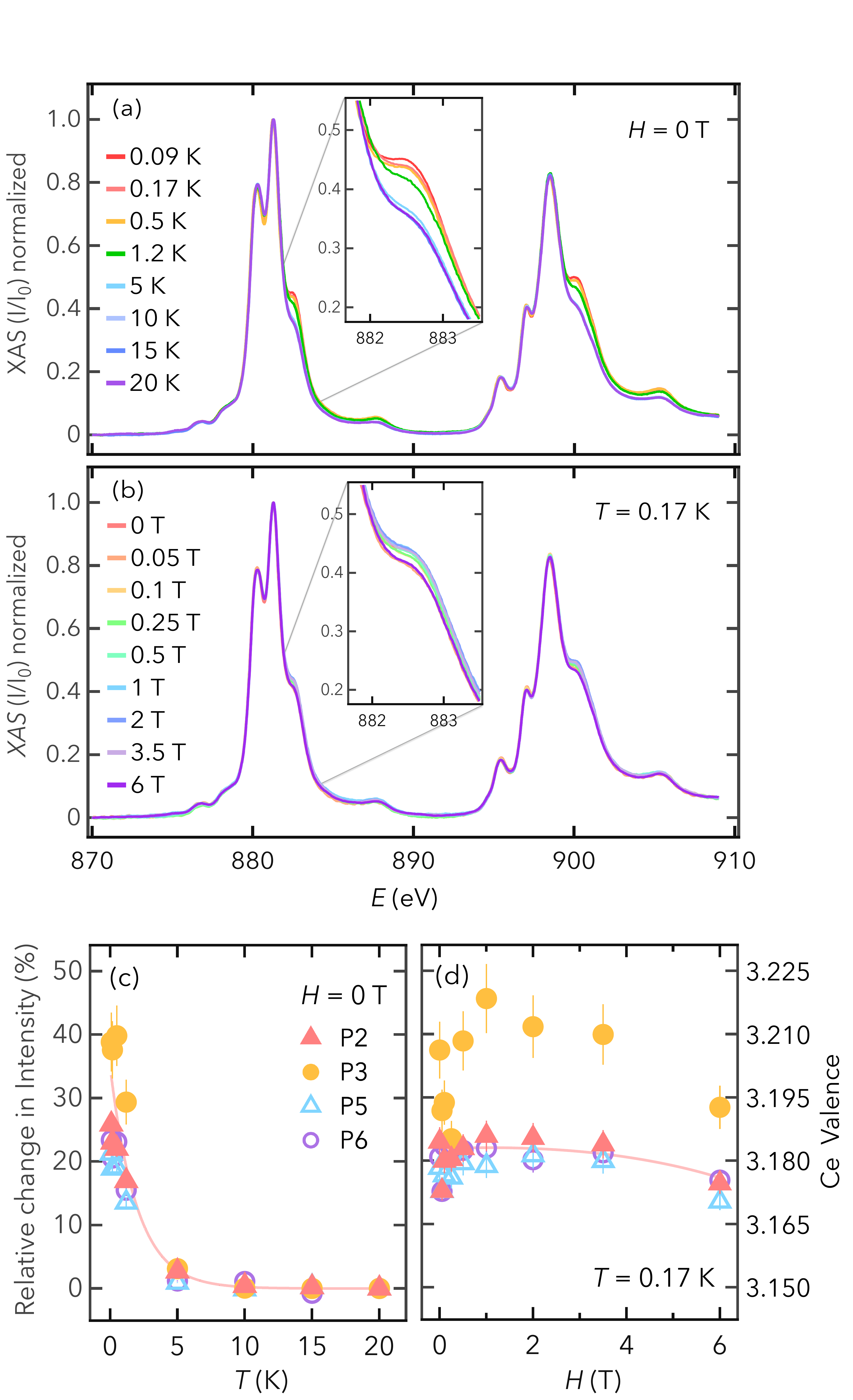}
\caption{\label{Fig3_Valence} (a) Isotropic Ce-M$_{5,4}$ XAS of CeCoIn$_5$ as a function of  temperature (normalized with respect to P1). (b) XAS as a function of applied external magnetic field. (c) Relative intensity change of the Ce$^{4+}$ absorption multiplets (Ce valence on right axis) as a function of temperature with the $20$\,K curve as reference. (d) Corresponding change as a function of magnetic field, using the same $20$\,K reference. Curves connecting data points in (c) \& (d) are guide to the eye.}
\end{figure}
Figure~\ref{Fig2_XAS} (b) shows the simulated XAS spectra for Ce$^{3+}$ and Ce$^{4+}$ in the atomic limit. Simulations were performed using CTM4XAS \cite{stavitski2010ctm4xas, thole19853d} with parameters from \cite{paidi2019role}. The M$_{5,4}$ peaks and shoulders are dominated by the $4\mathrm{f}^{1}$ and $4\mathrm{f}^{0}$ states. Both valences are clearly seen in the experimental data, indicating a mixed valence state. Contributions from Ce$^{4+}$ have been observed earlier \cite{howald2015evidence}, but appear to be more significant in this current study at very low temperatures.

Figure~\ref{Fig3_Valence} (a) shows the experimental XAS over the Ce M$_{5,4}$ edges at different low temperatures in zero magnetic field. Figure~\ref{Fig3_Valence} (b) shows the corresponding field-dependence at $0.17$\,K. The variation in $n_\mathrm{f}$ is determined by the relative intensity variations of the multiplets P2, P3, P5 and P6. We see a clearly observable change of relative intensities with temperature, but no pronounced change with magnetic field. Panels (c) and (d) of Fig.~\ref{Fig3_Valence}, show the temperature and magnetic field dependence of the relative intensity change of the Ce$^{4+}$ absorption multiplets with the $20$\,K curve as reference. This intensity change reflects change in $n_\mathrm{f}$ and the corresponding Ce valence is estimated on the right axis of panel (d) (see Fig.~S3 for valence evaluation). A pronounced increase in valence (decrease in $n_\mathrm{f}$) can be seen when the temperature is lowered. This increase in mixed valence state of Ce indicates an enhancement of charge fluctuations and reflects proximity to the QCP.
The Anderson impurity model (AIM) predicts that $n_\mathrm{f}$ should display properties with temperature dependence as a function of $T/T_\mathrm{K}$ \cite{bauer2004anderson}.
The increase seen in Fig.~\ref{Fig3_Valence} (c), however, occurs at very low temperature, below $5$\,K, much lower than the expected $T_\mathrm{K} \approx 45$\,K. We, therefore, argue that the increase is mainly driven by the approach to the QCP rather than Kondo effect alone. We do not see any significant field dependence at low temperature, see Fig.~\ref{Fig3_Valence}\,(d). This excludes superconductivity as the likely cause of the observed temperature dependence. However, at the $25^{\circ}$ external field orientation, the material stays in the superconducting state at $6$\,T with $\mu_0 H_{c2}\approx 8.8$\,T for the direction.

\begin{figure}
\includegraphics[width=0.98\linewidth]{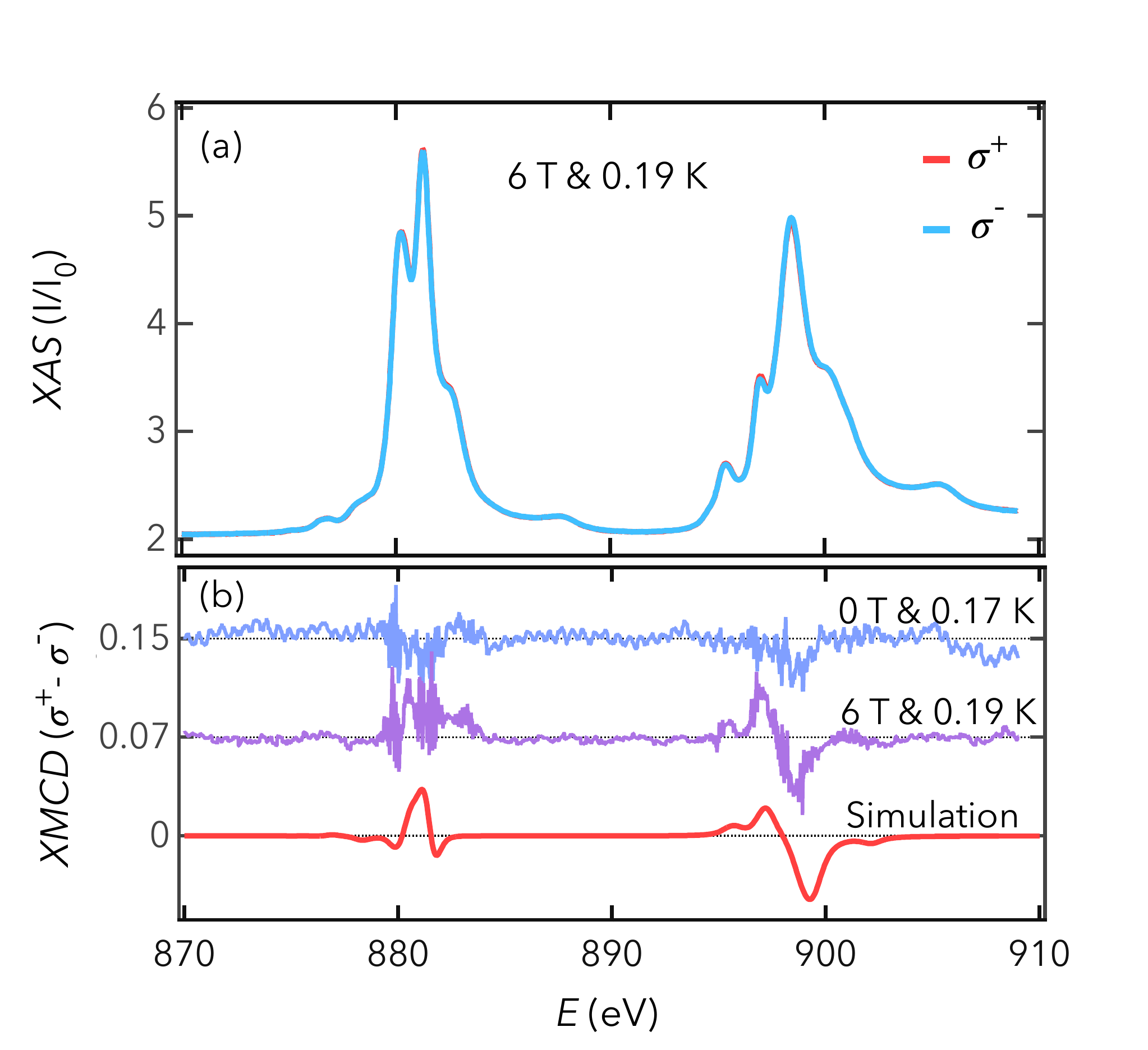}
\caption{\label{Fig4_XMCD}(a) Ce-XAS obtained by parallel ($\sigma^+$) and antiparallel ($\sigma^-$) circular polarized x-ray at 0.17\,K and 6\,T. (b) Simulated and experimental XMCD (0.17\,K \& 0.19\,K) for Ce at fields  0 \& 6\,T. XMCD is the difference of $\sigma^+$ and $\sigma^-$ XAS and matches with the simulated curve for experimental 6\,T, Baseline are shifted for visualization and are 0.150 and 0.075 for 0 \& 6\,T curve, respectively.}
\end{figure}

The effect of spin orientation can be probed by using circularly polarized XAS. The resulting XMCD, thus, can effectively be used to investigate the microscopic origin of magnetism at an elemental atomic level \cite{paidi2019role,gambardella2002ferromagnetism, schille19944f, thole1992x}. Figure~\ref{Fig4_XMCD} (a) shows a set of two oppositely circular polarized XAS spectra on the Ce M$_{5,4}$ edges measured in an applied magnetic field of $6$\,T at $0.19$\,K. The resulting XMCD contrast is shown in Fig.~\ref{Fig4_XMCD} (b) with XMCD at zero field for the same temperature. The Ce XMCD spectrum is only of the order of $\sim$1$\%$, but can nevertheless be measured reliably due to the good XAS signal to noise ratio. The XMCD signal becomes clear only at the highest fields ($6$\,T) in combination with low temperature.
We also studied the XAS and XMCD signal from Co (Fig.~S5). While the Co L$_{3,2}$ XAS displays pronounced multiplet structures resembling that of CoO (Co$^{2+}$) \cite{bora2015influence}, we find a negligible Co XMCD spectrum even for highest field of 6T at 0.19K (see supporting information). This is surprising because Co$^{2+}$ has a magnetic moment that cannot be completely quenched by CF and therefore should show some signature in the XMCD at high fields. Such signatures could be expected even with Kondo screening, since we probe the local spin state. The observations thus indicate collective  ion-ion interactions of antiferromagnetic character, such as in a spin liquid, for both Ce and Co.
To quantify the $\mathrm{Ce}^{3+}$ magnetic moment, XMCD spectra were simulated for the $[3\mathrm{d}^{10}]\ldots 4\mathrm{f}^{1} \rightarrow [3\mathrm{d}^{9}]\ldots 4\mathrm{f}^{2}$ transition in the atomic limit \cite{paidi2019role}. Interestingly, Ce M$_{5,4}$ XMCD spectral line shape is quite similar to the XMCD spectra of Co-CeO$_2$ and CeFe$_2$ ($J = 5/2$) \cite{paidi2019role,antonov2008x}, consistent with a still highly localized Ce moment. Comparison of simulations with experiments yields an average orbital angular momentum $\langle L_z \rangle =-0.324\,\hbar$ at the given conditions of Fig.~\ref{Fig4_XMCD}. Assuming that only the ground state $\ket{0}$ is occupied, the non-zero magnetic moment comes from magnetic field splitting of the ground state Kramers doublet into 
\begin{alignat}{2}
&\ket{0}_2 &&= 0.31 \ket{-5/2} + 0.95 \ket{+3/2}, \nonumber \\
&\ket{0}_1 &&= 0.31 \ket{+5/2} + 0.95 \ket{-3/2}. 
\end{alignat}
From the value of $\langle L_z \rangle$ we obtain a probability of $0.6$ for $\ket{0}_1$. This leads to $\langle S_z \rangle = 0.086\,\hbar$, $\langle L_z \rangle/\langle S_z \rangle = - 3.75$ and $\mu_\mathrm{avg} = 0.151\,\mu_B$ per Ce.

Under an applied external magnetic field, the Kramer's doublet can be described as a spin $m_s = \pm \hbar/2$ system \cite{alonso2015magnetic}. We write the Hamiltonian for the doublet as 
\begin{equation}
H = \mu_B \Vec{B} \cdot \Tilde{g}_\mathrm{eff} \cdot (\Vec{S}/\hbar),
\label{Eq:Hamiltonian}
\end{equation}
expressed in terms of an effective $\Tilde{g}_\mathrm{eff}$, expected to be $g_\mathrm{eff} = 1.4$ for isotropic conditions. In the absence of interactions, the population of levels would be controlled by temperature only. The observed level occupancies has a level splitting $37$ times smaller than that expected from Eq.~(\ref{Eq:Hamiltonian}) with $g_\mathrm{eff} = 1.4$ for the doublet. 

\begin{figure}
\includegraphics[width=0.98\linewidth]{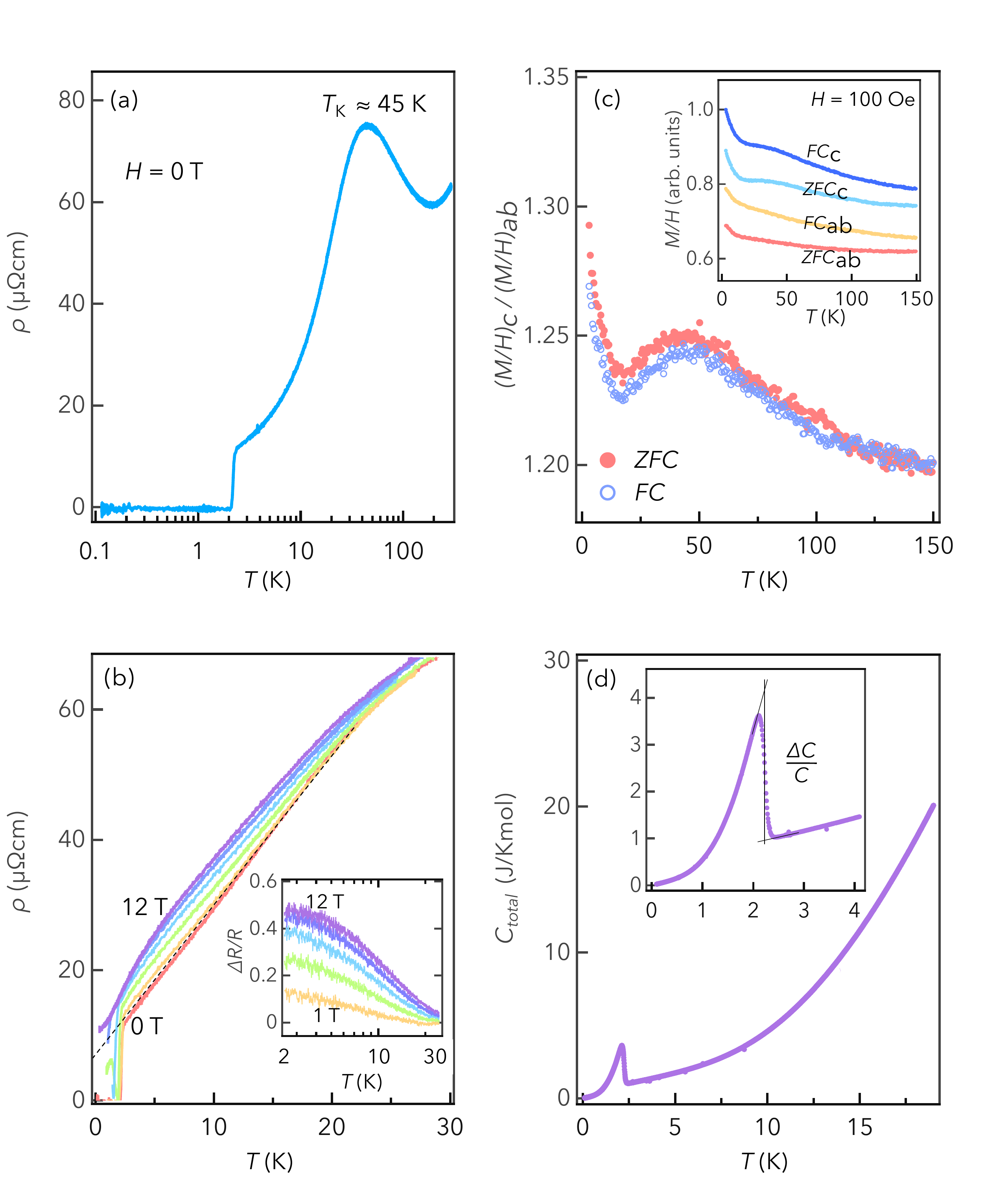}
\caption{\label{fig5}(a) In-plane resistivity of CeCoIn$_5$ as a function of temperature (T in log scale) in zero-field at a constant current ($0.3$\,mA) and frequency (23\,Hz). $T_\mathrm{K}$ corresponds to the Kondo transition and attains superconductivity at 2.3\,K, (b) In-plane resistivity of CeCoIn$_5$ in external fields parallel to the $ab$-plane. Inset shows the magnetoresistance for CeCoIn$_5$. (c) Magnetic anisotropy (M/H)$_c$/(M/H)$_{ab}$ at  zero field-cooled  (ZFC) and field cooled (FC) conditions, The inset shows ZFC and FC magnetisation of CeCoIn$_5$ as a function of temperature for two different crystallographic orientation in an applied field (H = 100\,Oe), the curves are separated with an offset, (d) shows the zero-field specific heat capacity as a function of temperature and a superconducting transition at 2.3\,K. the inset shows the unconventional superconducting jump in specific heat capacity.}
\end{figure}

The Kondo screening of magnetic moment on Ce and Co through conduction electrons can explain the negligible XMCD at zero field. For higher fields, however, we expect the Kondo screened magnetic spins to show up in XMCD at low temperature. CeCoIn$_5$ become superconducting at 2.3\,K as seen from resistivity, see Fig.~\ref{fig5} (a), but does not show signs of long range magnetic order in resistivity, magnetoresistance, magnetization, or specific heat, as shown in Fig.~\ref{fig5}.
The transition into a coherent Kondo scattering at $T_\mathrm{K} = 45$\,K as seen by in-plane resistivity Fig.~\ref{fig5} (a) relates with systematic changes of magnetoresistance below $T_\mathrm{K}$ shown in Fig~\ref{fig5} (b), indicating suppression of coherent scattering with magnetic field. CeCoIn$_5$ shows an approximately linear behavior below 20\,K in resistivity down to the superconducting transition. This strange metal behaviour is suppressed by the magnetic field. The inset of Fig.~\ref{fig5} (b) shows that the magnetoresistance increases with decreasing temperature and with increasing field parallel to the $ab$-plane. The rate of increase with field, however, slows at the highest fields.
The magnetic anisotropy of CeCoIn$_5$ is shown in Fig.~\ref{fig5} (c), taking the ratio of the magnetization of CeCoIn$_5$ at ZFC/FC conditions (inset) for the two different orientations. The effect of the Kondo transition is noticeably seen in the anisotropy, reflected mainly in the $c$-axis magnetization.
The unconventional character of superconductivity is seen in Fig.~\ref{fig5} (d) with specific heat as a function of temperature in zero field. The step of the superconducting transition, $\Delta C/C_e \geq 3.3$, is anomalously large for both conventional BCS ($\Delta C/C_e$ = 1.43) and strong-coupling superconductivity ($\Delta C/C_e$ reaching about 2.5) \cite{johnston2013elaboration, kuzmenko2020bcs}.

Without signs of long-range order, a short-range antiferromagnetic pairing mechanism is required to explain the weak XMCD signal for Ce and negligible signal for Co. Such spin-liquid-like behavior, as described in Fig.~\ref{Fig1_QCP_phase_diagram}, seems to be present both for Ce and Co. Cerium, however, has a temperature dependent valence which increases as the temperature is decreased. These valence fluctuations may cause the local AFM pairing to weaken. Furthermore, if the valence fluctuations are allowed to move in the lattice, charge-density-waves (CDW) would arise.

In summary, we find that Ce in CeCoIn$_5$ has a temperature dependent mixed valence state at low temperature. This suggests that CeCoIn$_5$ is intrinsically close to a QCP with associated valence fluctuations. The XMCD shows a weak signal for the Ce ground state Kramers doublet in magnetic field indicating antiferromagnetic spin-liquid-like interactions. The significant low-temperature change in Ce valence may promote charge fluctuations coexisting with the spin fluctuations, possibly involved in the microscopic origin of unconventional superconductivity in CeCoIn$_5$.

\section{acknowledgments}
This project work was partially supported through DST project IIT BHU/DST/R\&D/SMST/09. A. Khansili and A. Rydh acknowledges support by the Swedish Research Foundation, D. Nr. 2021-04360. We thank A. Thamizhavel for providing the crystals and P. D. Babu, S. Thirupathiah for their extended help. We acknowledge DESY (Hamburg, Germany), a member of the Helmholtz Association HGF, for the provision of experimental facilities. Parts of this research were carried out at P04 PETRA II and beamtime was allocated for proposal I-20200389 under the India-DESY collaboration program.

\section{Author contributions}
S. K. Mishra, A. Khansili, and A. Rydh conceived the project, designed and planned all the experiments. The magnetic measurements were performed by A. Khansili, P. D. Babu, and R. Hissariya. XAS have been carried out at DESY P04 beamline by I. Baev, J. Schwarz, F. Kielgast, M. Nissen, M. Martins, M.-J. Huang, and M. Hoesch. Data analysis and simulation has been done by A. Khansili, and R. Sharma with the contribution of S. K. Mishra, A. Rydh, V. K. Paidi and J.v. Lierop. The transport measurements were conducted by A. Khansili and A. Rydh. The manuscript was drafted by A. Khansili, R. Sharma, A. Rydh, and S. K. Mishra with contributions from all coauthors.

\appendix

\end{document}


\title{Supplemental material: Element-specific probe of quantum criticality in CeCoIn$_5$}

\author{A. Khansili}
\affiliation{School of Materials Science and Technology, Indian Institute of Technology (Banaras Hindu University), Varanasi -  221005, India}%
\affiliation{Department of Physics, Stockholm University, SE-106 91 Stockholm, Sweden}%

\author{R. Sharma}
\affiliation{School of Materials Science and Technology, Indian Institute of Technology (Banaras Hindu University), Varanasi -  221005, India}%
 
\author{R. Hissariya}
\affiliation{School of Materials Science and Technology, Indian Institute of Technology (Banaras Hindu University), Varanasi -  221005, India}%

\author{I. Baev}
\affiliation{Universit{\"{a}}t Hamburg, Institut f{\"{u}}r Experimentalphysik Luruper Chaussee 149, Hamburg, Germany}%

\author{J. Schwarz}
\affiliation{Universit{\"{a}}t Hamburg, Institut f{\"{u}}r Experimentalphysik Luruper Chaussee 149, Hamburg, Germany}%

\author{F. Kielgast}
\affiliation{Universit{\"{a}}t Hamburg, Institut f{\"{u}}r Experimentalphysik Luruper Chaussee 149, Hamburg, Germany}%

\author{M. Nissen}
\affiliation{Universit{\"{a}}t Hamburg, Institut f{\"{u}}r Experimentalphysik Luruper Chaussee 149, Hamburg, Germany}%

\author{M. Martins}
\affiliation{Universit{\"{a}}t Hamburg, Institut f{\"{u}}r Experimentalphysik Luruper Chaussee 149, Hamburg, Germany}%

\author{M. -J. Huang}
\affiliation{Deutsches Elektronen-Sychrotron DESY, Notkestraße 85, 22607 Hamburg, Germany}%

\author{M. Hoesch}
\affiliation{Deutsches Elektronen-Sychrotron DESY, Notkestraße 85, 22607 Hamburg, Germany}%

\author{V. K. Paidi}
\affiliation{Pohang Accelerator Laboratory, Pohang 37673, South Korea}%
\affiliation{Department of Physics and Astronomy, University of Manitoba, Winnipeg, Manitoba R3T 2N2, Canada}%

\author{J. van Lierop}
\affiliation{Department of Physics and Astronomy, University of Manitoba, Winnipeg, Manitoba R3T 2N2, Canada}%

\author{A. Rydh}
\email{andreas.rydh@fysik.su.se}
\affiliation{Department of Physics, Stockholm University, SE-106 91 Stockholm, Sweden}%

\author{S. K. Mishra}
\email{shrawan.mst@iitbhu.ac.in}
\affiliation{School of Materials Science and Technology, Indian Institute of Technology (Banaras Hindu University), Varanasi -  221005, India}%

\maketitle

\section*{EDX}
\begin{figure}[ht]
\includegraphics[width=0.5\linewidth]{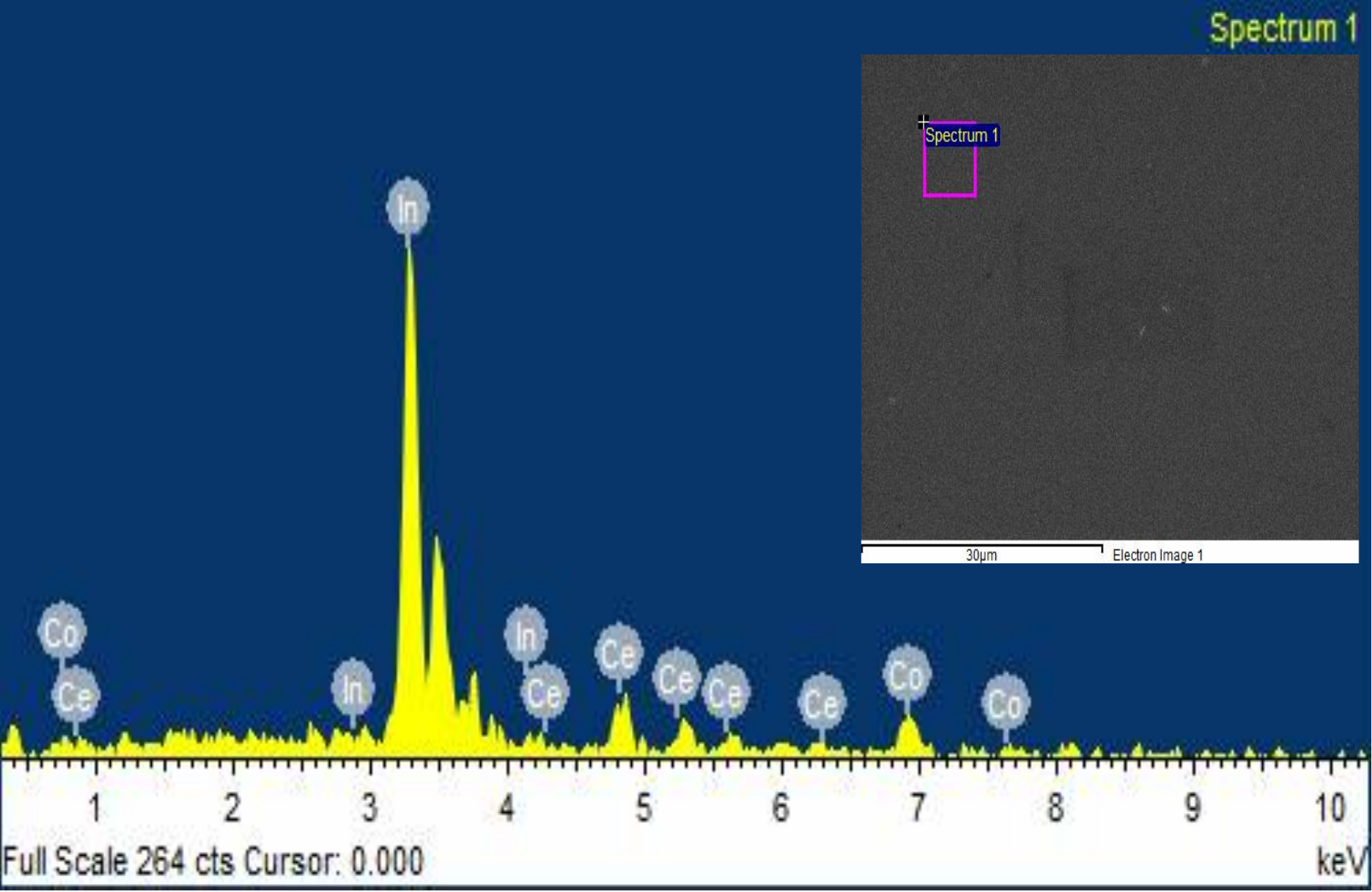}
\caption{\label{SI1}
Compositional analysis using EDX. The inset of the figure shows a SEM image of the analyzed surface.}
\end{figure}

The composition of CeCoIn$_5$ was examined through scanning electron microscope (SEM) with energy-dispersive x-ray spectroscopy (EDX). Figure~\ref{SI1} shows the chemical composition and microscopic surface of a crystal.  The measurement was performed at various locations in the crystal surface to confirm the suitable composition and homogeneity. The average chemical composition, shown in Table~\ref{tab:table1}, is consistent with the expected composition for CeCoIn$_5$ (Ce: Co: In = 1: 1: 5) within uncertainties of the EDX technique.
\begin{table}[h]
\centering
\begin{tabular}{c@{\hspace{2cm}}c}
\hline
    Element & Atomic\,\%  \\
\hline
    Cerium & 13.65  \\ 
    Cobalt & 15.16  \\ 
    Indium & 71.19  \\ 
\hline
\end{tabular}
    \caption{Compositional data from EDX.}
    \label{tab:table1}
\end{table}

\section*{Magnetization loops}
\begin{figure}[h]
\includegraphics[width=0.55\linewidth]{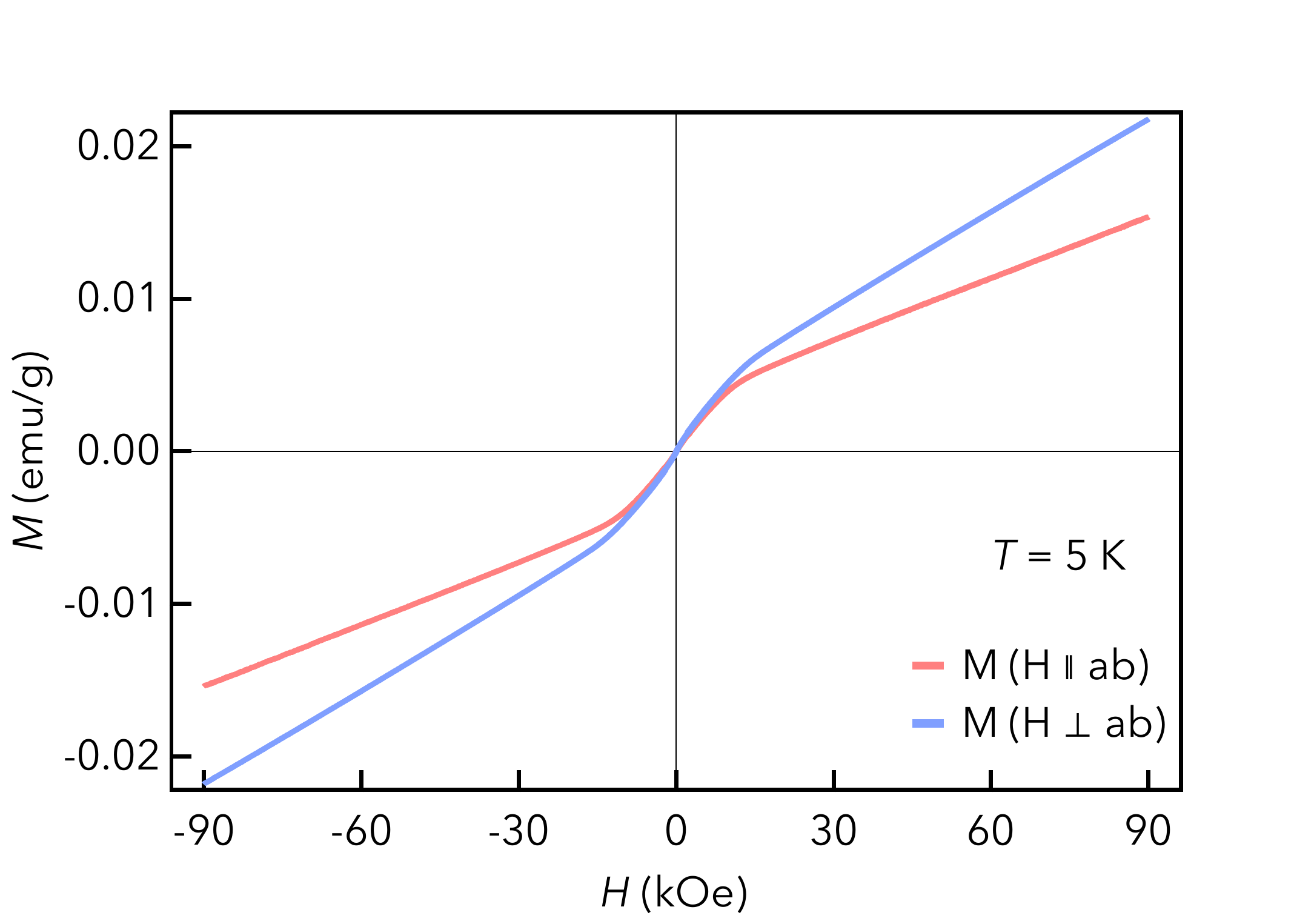}
\caption{\label{MH_loop} The M-H loop can be seen in the figure for a bipolar 9\,T external field in two different orientations for the single crystal at 5\,K temperature.}
\end{figure}

Figure~\ref{MH_loop} shows a $M$-$H$ loop for CeCoIn$_5$ at $5$\,K in applied magnetic field up to 9\,T for two different crystallographic orientations.
\begin{figure}[h]
\includegraphics[width=0.95\linewidth]{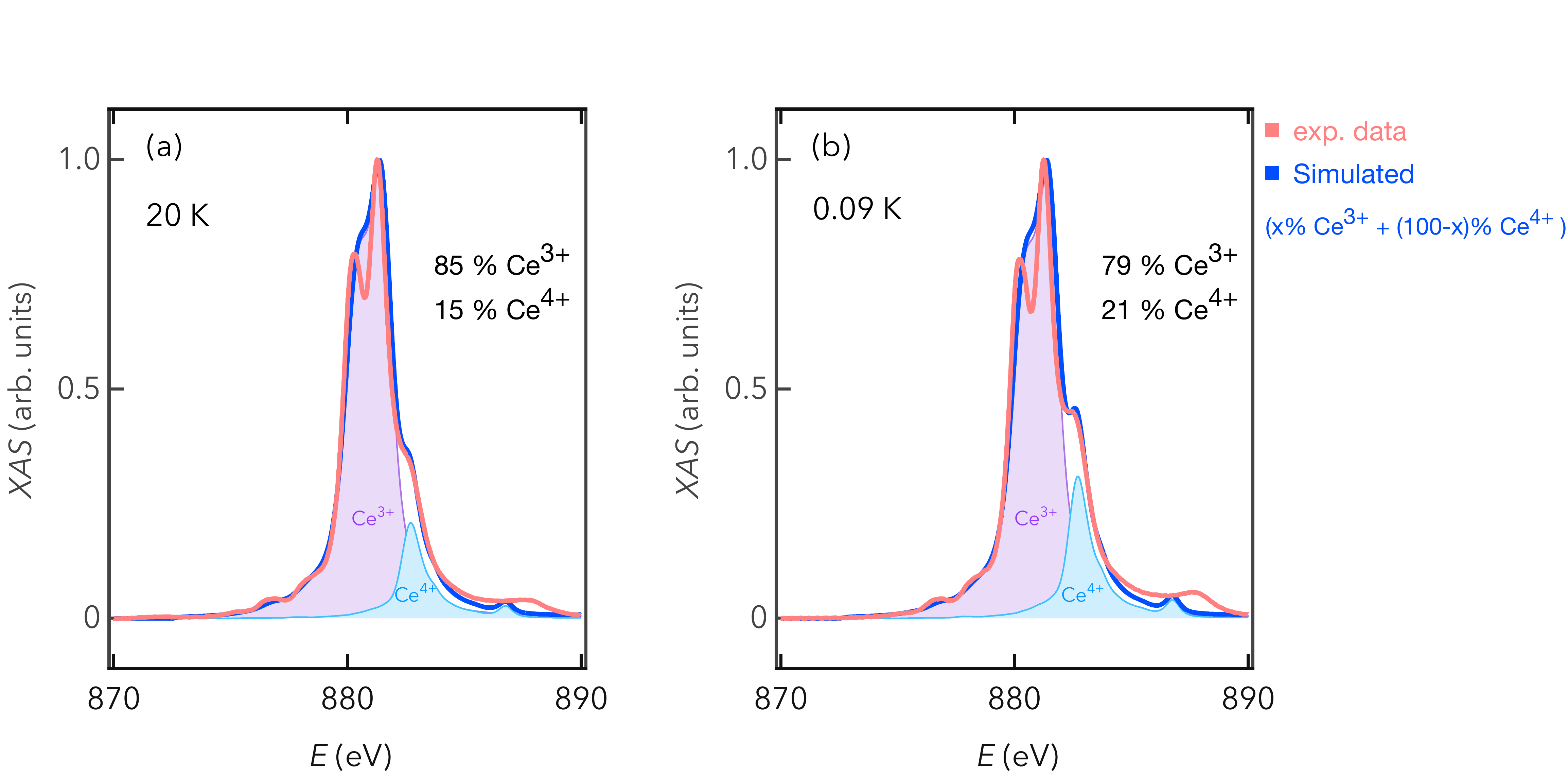}
\caption{\label{Valence_Fit} (a) Ce-XAS for CeCoIn$_5$ at M$_{5}$ absorption edge at 20\,K with simulated data shown in the same graph. Simulated curve at 20\,K has 85\% Ce$^{3+}$ and 15\% Ce$^{4+}$ contribution which is the best estimate for the given simulations. (b) Ce-XAS for CeCoIn$_5$ at M$_{5}$ absorption edge at 0.09\,K with simulated data. Simulated curve at 0.09\,K shows an increased contribution of 21\% Ce$^{4+}$ from 15\% at 20\,K.}
\end{figure}
\section*{Ce Valence Evaluation}
Figure~\ref{Valence_Fit}(a \& b) illustrate the normalized M$_{5}$ edge experimental and simulated data with contributions from Ce$^{3+}$ \& Ce$^{4+}$ for 20\,K and 0.09\,K respectively. M$_5$ edge is suitable for fitting due to the good agreement with the simulated data. There are additional small features on the M$_{5,4}$ edges that are not seen in the simulated data. However, these features do not have a temperature and magnetic field dependence. The significant peaks for the Ce$^{4+}$ have a clear temperature dependence. At 20\,K, Ce has a mixed valence of +3.15 with 15\% of Ce$^{4+}$. This fraction is increased as the temperature is decreased to 21\% Ce$^{4+}$ (Ce valence +3.21) at base temperature.
\begin{figure}[h]
\includegraphics[width=0.95\linewidth]{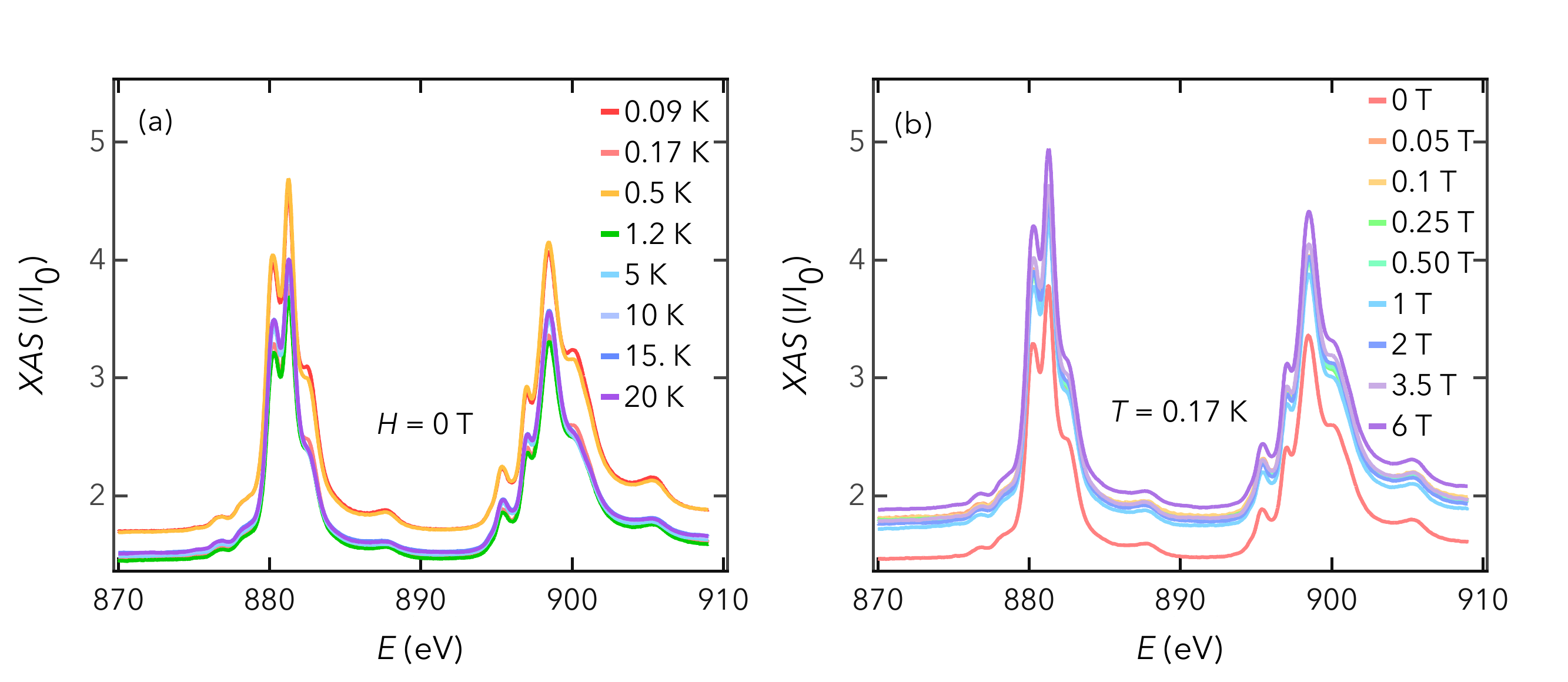}
\caption{\label{XAS_Raw} (a \& b) Raw Ce-XAS for CeCoIn$_5$ at M$_{5,4}$ absorption edges of cerium as a function of external magnetic field and temperature.}
\end{figure}
Figure~\ref{XAS_Raw}(a \& b) shows the isotropic raw XAS data for Ce as a function of temperature and external field. I is the total electron yield (TEY) signal and I$_0$ is the incoming beam intensity measured for each XAS scan individually. For each XAS curve, we measure two different oppositely polarized XAS data. Each polarized condition includes 8 separate scans and are averaged to reduce the noise in the data. To get the isotropic XAS, we average the positive and negative polarized XAS. The intensity plotted  (I/I$_0$) has different background and peak intensity, however, the shape of the M$_{5,4}$ absorption edges remains the same and can be compared by subtracting the background for each scan (keeping pre-edge at 0) and then normalizing to the M$_5$ peak to analyse the change in the Ce$^{4+}$ state.\\ 

The Ce XAS could potentially include an additional f$^2 \rightarrow$ f$^3$ contribution. Such additional contribution from Ce$^{2+}$ (f$^2 \rightarrow$ f$^3$) is expected at higher energies compared to f$^1 \rightarrow$ f$^2$ \cite{praetorius2015m4} with a similar lineshape as for Pr\cite{thole19853d}. However, unlike the f$^0 \rightarrow$ f$^1$ contribution, f$^2 \rightarrow$ f$^3$ contribution does display XMCD. Since we do not observe any XMCD for peaks P2, P3, P5 and P6 (see main text Fig.~3\,$\&$\,4), we unambiguously conclude that these are from Ce$^{4+}$ (f$^0 \rightarrow$ f$^1$).
\begin{figure}
\includegraphics[width=0.95\linewidth]{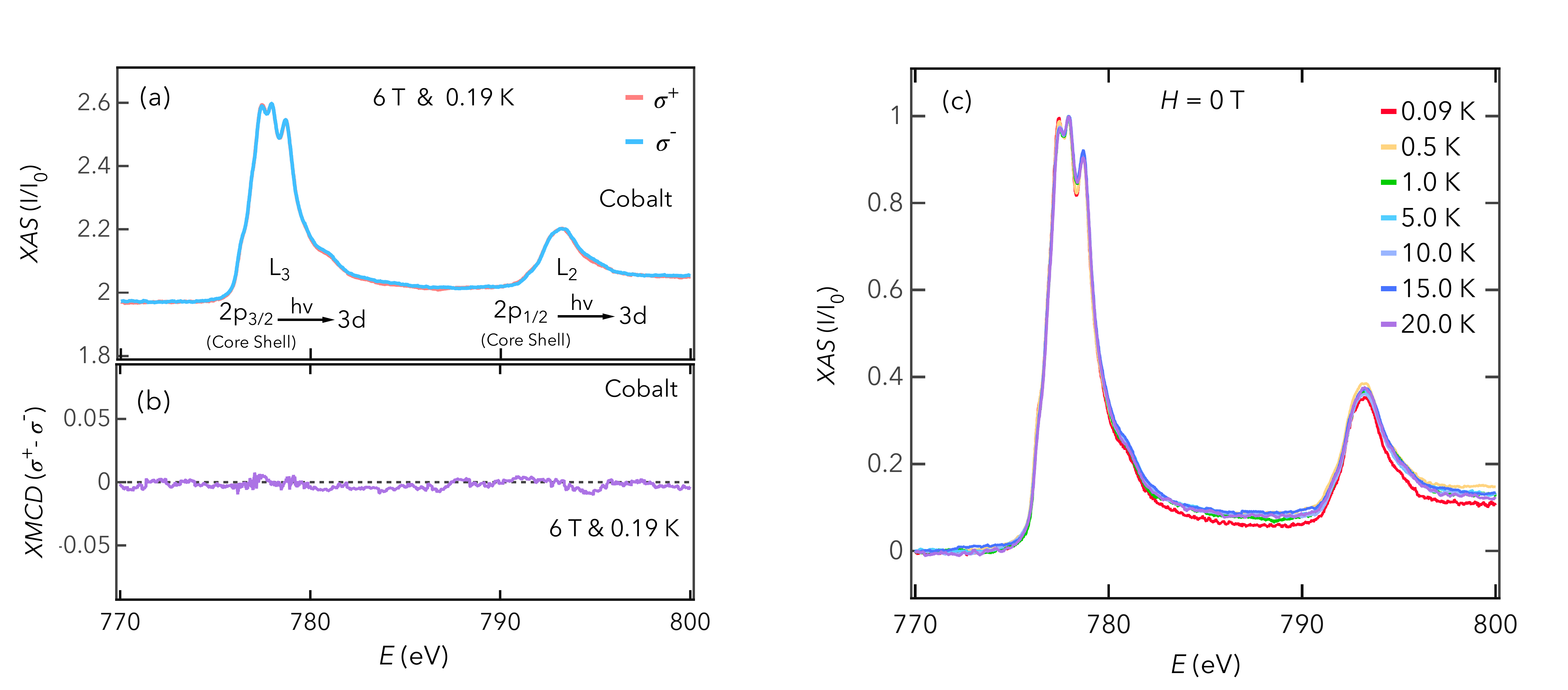}
\caption{\label{Co_XAS_XMCD} (a) Co-XAS obtained by parallel ($\sigma^+$) and antiparallel ($\sigma^-$) circular polarized x-ray at 0.19\,K and 6\,T. (b) Co-XMCD at 0.19\,K and 6\,T. (c) Temperature dependence of normalized isotropic XAS of Co at zero external magnetic field.}
\end{figure}
\section*{Co XAS $\&$ XMCD}
Figure~\ref{Co_XAS_XMCD}(a) shows the XAS of Cobalt at 6\,T and 0.19\,K. The multiplet structure resembles that of CoO\cite{bora2015influence} with cobalt dominated by Co$^{2+}$ oxidation states. Figure~\ref{Co_XAS_XMCD}(b) shows the negligible XMCD contrast even at 6\,T at 0.19\,K. Figure \ref{Co_XAS_XMCD}(c) shows the temperature dependence of Co-XAS in the absence of external magnetic field. The XAS is normalized to the second multiplet feature of the first (L$_3$) peak. There is no aparent temperature dependence of the XAS spectra for Co. There seems to be negligible XMCD for Co at all measured fields and temperatures. The absence of XMCD signal at high fields, in addition with the lack of long range order suggests that Co has a local antiferromagnetic pairing interaction (main text Fig~1). A strong Kondo screening can completely screen the magnetic moment on Co at zero field. However, the Co moment would still be affected by the applied external magnetic field and would be detectable at high field ($\sim$6\,T). Since we do not observe any XMCD for Co at 6\,T and 0.19\,K, it would be hard to explain the results with only Kondo screening of Co ions.
